\documentclass[pra,twocolumn,showpacs]{revtex4}

\usepackage{bm}
\usepackage{graphicx}
\usepackage{amsmath}
\usepackage{subeqnarray}

\newcommand{\n}{\nonumber}
\newcommand{\bn}{\begin{eqnarray}}
\newcommand{\en}{\end{eqnarray}}
\newcommand{\h}{\hspace}

\begin{document}

\title {Conditions for Vanishing Central-well Population in Triple-well Adiabatic Transport}
 \author{Tom\'{a}\v{s} Opatrn\'{y}$^1$ and Kunal K. Das$^{2,3}$}
 \affiliation{$^1$Optics Department, Faculty of Science, Palack\'{y} University, 17. Listopadu 50,
 77200 Olomouc, Czech Republic}
\affiliation{$^2$Department of Physical Sciences, Kutztown University of Pennsylvania, Kutztown, PA 19530, USA}
\affiliation{$^3$Department of Physics, Fordham University, Bronx, New York 10458, USA}

\date{\today }
\begin{abstract}
Analytical expressions are derived for coherent tunneling via adiabatic passage (CTAP) in a triple well system with negligible central-well population at all times during the transfer.  It is shown that a manipulation of the depths of the extreme-wells, correlated with the time variation of the \emph{non-adjacent} barriers is essential for maintaining vanishing population of the central well. The validity of our conditions are demonstrated with a numerical solution of the time-dependent Schr\"{o}dinger equation. The transfer process is interpreted in terms of a current through the central well.
\end{abstract}

 \maketitle

\section{Introduction}

\begin{figure}[b]
\centering
\includegraphics[width=0.7\linewidth]{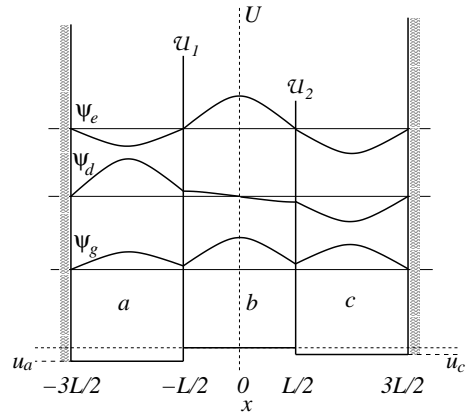}
\caption{Potential structure for the three-well system with schematic plots of the three lowest energy eigenstates. The three wells $a$, $b$,
and $c$ of length $L$ are separated by two delta-function barriers of strengths ${\cal U}_{1,2}$. The potential of the middle well is set to 0, the outer wells $a$ and $c$ have variable potentials $U_{a,c}$.
\label{fig1}}
\end{figure}

Certain characters in the Harry Potter adventures have the magical power to move from one room to another without having to go through the staircase in between \cite{Rowling}.  Even without resorting to magic and wizardry, something quite similar can be brought to pass in the quantum world: among three adjacent potential wells $a$, $b$, and $c$ (see Fig. \ref{fig1}),  a particle in well $a$ at one end can be transferred to the non-adjacent well $c$ at the other end without bypassing or occupying the well $b$, that lies in between. The principle behind such a transfer has been well established for quite
some time in the context of internal atomic states for a two-photon process known as STIRAP (stimulated Raman adiabatic pumping) \cite{STIRAP}, wherein atoms are pumped from a state  $a$ in their electronic ground-state manifold to another state  $c$ in the same manifold by means of a pair of time varying laser fields that couple  $a$ and $c$ to an excited state  $b$.

The idea can be applied to other similar manifestations of discrete quantum states; specifically when those states represent the spatial wavefunctions of physical particles, one can achieve the seemingly magical transport between two locations without traversing the intervening space.  The last few years have witnessed a surge of theoretical interest in applying the STIRAP principle to spatial transport of physical particles in a variety of systems:  atoms in optical microtraps \cite{Eckert}, electrons in quantum dots \cite{Greentree, Cole}, Bose-Einstein condensates \cite{Graefe,Rab}, superconductors \cite{Siewert-superconductors}, spin transport in the context of quantum information \cite{Hollenberg-2006}.  Experimental demonstration was achieved for photons in optical waveguides \cite{Laporta-PRB}. Even a new acronym CTAP (coherent tunneling via Adiabatic Passage) has been coined to describe the spatial analog \cite{Greentree}. The popular theoretical description of the process
is borrowed from quantum optics, and is in terms of a discrete $3\times 3$  Hamiltonian matrix with variable off-diagonal elements corresponding to the coupling between the traps. Here, we adopt a more flexible approach, appropriate for spatial transfer, where the evolution is described by a time dependent Schr\"{o}dinger equation for the appropriate potentials forming the wells and the barriers.  The advantage of such a method lies in its capability to model arbitrary potential shapes and variations accurately \cite{Cole,Rab}.

Regardless of the method used, results on such adiabatic transfer have been  numerical in nature, with the tradeoff that the parametric dependencies
of the transfer process are hard to identify. Our goal in this paper is to obtain simple analytical conditions for the system parameters by considering a somewhat simplified triple-well model which captures the essential features of more complex potentials. We find that in order to have a vanishing population of the  central well at all times certain supplementary conditions need to be satisfied: First, out of the three lowest-energy and relevant eigenstates of the Hamiltonian, the middle-energy one should maintain as low a population of well $b$ as possible, which can be achieved by requiring its node to be at the center of well $b$. Secondly the energy eigenvalues of the instantaneous Hamiltonian need to be manipulated in such a way that the process stays in the adiabatic regime, which can be achieved by keeping the energy of the middle state as far from the adjacent states as possible. We show that the implementation of these conditions necessitate the manipulation of the well depths as well as the barrier heights.

The paper is structured as follows: In Sec. II, we derive the conditions for keeping the central well population negligible at all times in the limiting case of delta-function potential barriers, and in Sec. III we show that those conditions apply for finite width barriers that are narrow and high at all times, and also qualitatively for wider barriers. These conditions are applied in a numerical simulation based on a time-dependent Schr\"{o}dinger equation in Sec. IV to demonstrate the validity of our conditions. In Sec. VI, we provide an interpretation of the transfer process in terms of a high velocity current, and summarize our results in the conclusions section.


\section{System parameters and eigenstates}

In STIRAP the transfer is achieved by using a counterintuitive coupling sequence, where $b$ is first coupled to the empty target state $c$, and only when that coupling is reduced, the $ab$ coupling is ramped up. Appropriate adiabatic variation of the two couplings map the state $b$ to a dark state superposition of states  $a$ and $c$, which contains no contribution of state $b$. The name `dark state' refers to the fact that atoms in this superposition cannot
interact with the laser fields. By gradually switching off the $ab$ coupling, the dark state is adiabatically transformed into  $c$. During the process, there is virtually no occupation of the intermediate state $b$ which would spontaneously decay into one of lower energy states.

In the spatial analog, the three states are implemented by three adjacent potential wells with the \emph{inverse} strengths of the two intra-well barriers signifying the coupling between the wells.  We implement this process by considering a one dimensional system with the potential energy structure as follows (see Fig. \ref{fig1}):
\begin{eqnarray}
V(x)={\cal U}_1 \delta\! \left(x+{\textstyle\frac{L}{2}}\right)
+ {\cal U}_2 \delta\! \left(x-{\textstyle\frac{L}{2}}\right)
+\left\{
 \begin{array}{lc}
  \infty, & x\le -\frac{3}{2}L \\
  U_a, & x\in a \\
  0, & x\in b \\
  U_c, & x\in c\\
 \infty, &  \frac{3}{2}L \le x
 \end{array}
 \right.
\end{eqnarray}
where each well spans a length $L$ in the intervals $
a~\equiv~\left[-\frac{3}{2}L ; -\frac{1}{2}L \right],
b\equiv\left[ -\frac{1}{2}L ; \frac{1}{2}L\right],
 c\equiv\left[ \frac{1}{2}L ; \frac{3}{2}L\right]$.
Effective 1D behavior can be implemented for atoms in micro-magnetic traps
\cite{atom-chip} with tight transverse confinement, and for electrons in size-quantized nanowires \cite{nanowire}.  The delta-potentials are idealizations of narrow and high potential barriers, which can be implemented with tightly focused lasers for ultracold atoms, or sharp field-gradient between adjacent quantum dots in the case of electrons.  The potential energy shifts $U_{a,c}$ are analogs of detuning among internal atomic states \cite{Nesterenko}. Denoting the fundamental wavevector for an infinite square well of length $L$ by $k_0\equiv \pi/L$ for a state with wavelength $\lambda=2L$, we define the dimensionless coupling parameters $\alpha_{1,2}$ for the \emph{inverse}-strengths of delta-function potentials, and
$u_{a,c}$ for the constant energy shifts of the two outer wells:
\begin{eqnarray}
\label{eq-ualpha}
 {\cal U}_{1,2} = \frac{\hbar^2 k_0}{2\pi m \alpha_{1,2}} \h{1cm}
 U_{a,c} = \frac{\hbar^2 k_0^2u_{a,c}}{2m} .
\end{eqnarray}
Consistent with the assumption of high barriers, we assume $\alpha_{1,2} \ll 1$ during the whole process so that only the lowest orders of $\alpha_{1,2}$ need to be considered.

The initial conditions, along with the adiabaticity, can ensure that only the three lowest energy states of the instantaneous Hamiltonian are energetically accessible and relevant; we denote the states by $\psi_g$ (ground), $\psi_d$ (dark), and $\psi_e$ (excited), in order of increasing energy. We study their properties below to extract conditions on the system parameters.

\subsection{Dark state}

The middle-energy eigenstate plays the crucial role in the CTAP time evolution, by evolving as a dark state with little contribution from the central well, b.  In order to minimize the population of well $b$, we assume a symmetric form of the middle-energy eigenstate
\begin{eqnarray}
 \label{eq-psi-1}
 \psi_d(x)=\left\{
 \begin{array}{lc}
 A_d \sqrt{\frac{2}{L}}
 \cos \left[k_0(1-y_a)x- \frac{3\pi}{2}y_a \right], & x\in a \\ 
  B_d\sqrt{\frac{2}{L}}\sin  \left[k_0(1-y_b)x \right], & x\in b \\ 
 C_d\sqrt{\frac{2}{L}}\cos \left[k_0(1-y_c)x+ \frac{3\pi}{2}y_c \right], & x\in c\\ 
\end{array}
 \right.
\end{eqnarray}
The small parameters $y_{a,b,c} \ll 1$ correspond to phase shifts of the infinite well wavevector due to the finite heights and widths of the intermediate barriers. The conditions of continuity of $\psi_d$ at $x=\pm L/2$, on retaining only terms linear in  $y_{a,b,c}$ lead to the relations
\begin{eqnarray}
 \label{eq-abc-y}
 A_d\pi y_a = B_d= -C_d\pi y_c .
\end{eqnarray}
The time-independent Schr\"{o}dinger equation applied at the positions of the  delta walls leads to the requirements
\begin{eqnarray}
 \label{eq-yaq}
 y_{a,c} &=& \alpha_{1,2},
\end{eqnarray}
and applied within the wells, gives
\begin{eqnarray}
\frac{2m E}{\hbar^2 k_0^2}=(1-y_a)^2 + u_a= (1-y_b)^2
 = (1-y_c)^2 + u_c ,
\end{eqnarray}
from which it follows that
\begin{eqnarray}\label{energy-condition}
 y_a - \frac{u_a}{2} = y_b = y_c - \frac{u_c}{2}
\end{eqnarray}
up to first order in $y_{a,b,c}$.
This leads to the condition connecting the potentials of the walls to the inverse barrier strengths:
\begin{eqnarray}
\label{eq-ucua}
 u_c-u_a = 2(\alpha_1-\alpha_2).
\end{eqnarray}
This relation underscores the fact that slight shifts in the potential wells are essential to minimize the population of the central well, and the optimal shifts need to be varied synchronously with the barrier potentials between the wells.  If this condition is not satisfied then the dark state does not have the node in the middle of the central well, instead the node is closer to (or inside of)
one or the other of the two walls, which due the heightened asymmetry, increases the net probability of the particle to be in the central well. Such, for instance would be the case in the model used in Ref.~\cite{Cole} where the extreme wells have equal depths, which in our notation would mean $u_c=u_a$.

The normalization coefficients in (\ref{eq-psi-1}) are determined by Eqs. (\ref{eq-abc-y}) and (\ref{eq-yaq}) to be
\begin{eqnarray}
 A_d\!= \!\frac{-\alpha_2}{\sqrt{\alpha_1^2 + \alpha_2^2}},\
 B_d \!=\!\frac{-\pi \alpha_1 \alpha_2}{\sqrt{\alpha_1^2 + \alpha_2^2}}, \
 C_d \!=\! \frac{\alpha_1}{\sqrt{\alpha_1^2 + \alpha_2^2}},
\end{eqnarray}
on keeping the lowest order terms in $\alpha_{1,2}$, and the normalization $|A_d|^2+|B_d|^2+|C_d|^2=1$ is satisfied up to the same order. This assumes that the barrier heights are always kept high, forcing the parameters $\alpha_{1,2}$ to be always small. In that case, the population of the middle well, $|B_d|^2$ can be made arbitrary small compared to the other two wells, since the numerator of $B$ is bilinear in the small parameters, but the numerators of $A_d$ and $C_d$ are linear, while they all share the same denominator which is effectively of linear order in the small parameters.  It is clear from the expressions that by changing the ratio of the magnitudes of the barriers and therefore $U_1:U_2=\alpha_2:\alpha_1$ the population ratio of wells $a$ and $c$ can vary as the squares, $\alpha_2^2:\alpha_1^2$, from $1:0$ to $0:1$.

\subsection{Ground and excited states}

The condition Eq.~(\ref{eq-ucua}) fixes an optimal difference for the extreme well depths, but leaves the actual values free. Further optimization will require a knowledge of the remaining states. The ground ($g$) and the second excited ($e$) states of the Hamiltonian both have the general form
\begin{eqnarray}
 \label{eq-psi-2}
 \psi_{g,e}(x)=\left\{
 \begin{array}{lc}
A \sqrt{\frac{2}{L}}\cos \left[k_0(1-y_a)x- \frac{3\pi}{2}y_a \right], &  x\in a \\
B \sqrt{\frac{2}{L}}\cos  \left[k_0(1-y_b)x + \phi\right], &  x\in b \\
C \sqrt{\frac{2}{L}}\cos \left[k_0(1-y_c)x+ \frac{3\pi}{2}y_c \right], &  x\in c
\end{array}
 \right.
\end{eqnarray}
where we assume $y_{a,b,c} \ll 1$ and $\phi \ll 1$. The  continuity of $\psi_{g,e}$ at $x=\pm L/2$ leads to the conditions
\begin{eqnarray}
 \label{eq-abc-phi1}
 A\pi y_a =-B \left( \frac{\pi}{2}y_b +\phi \right);\h{4mm}
 C\pi y_a = -B \left( \frac{\pi}{2}y_b -\phi \right).
\end{eqnarray}
The stationary Schr\"{o}dinger equation applied at the position of the delta walls leads to the requirements
\begin{eqnarray}
 \label{eq-abc-qya}
 A\left(1-\frac{y_a}{\alpha_1}\right) = B =  C\left(1- \frac{y_c}{\alpha_2}\right),
\label{eq-abc-qyb}
\end{eqnarray}
while energy eigenvalues in the wells lead to a condition identical to  Eq.~(\ref{energy-condition}) for the dark state.  Equations (\ref{eq-abc-phi1}) yield the following relations
\begin{subeqnarray}
 \phi= \frac{\pi}{2b}(C y_c - A y_a),\\
 A y_a + B y_b + C y_c =0.
\end{subeqnarray}
Thus all the parameters of the wavefunction are determined by the shift $y_b$ of the wavenumber in the middle well, which we now solve for after simplifying the notation by dropping the subscript: $y\equiv y_b$. Expressing $y_{a,c}$ by means of
Eq.~(\ref{energy-condition}), and $A,C$ by means of Eqs. (\ref{eq-abc-phi1})-(\ref{eq-abc-qyb}) we find the requirement for $y$ to be
\begin{eqnarray}
\label{eq-y-1}
  \frac{\alpha_1\left( y+\frac{u_a}{2} \right)}{\alpha_1- y-\frac{u_a}{2} }
+y + \frac{\alpha_2\left( y+\frac{u_c}{2} \right)}{\alpha_2- y-\frac{u_c}{2} }=0.
\end{eqnarray}
Imposing Eq. (\ref{eq-ucua}) for minimum middle well population, the potential shifts of the well can be written as
\begin{eqnarray}
u_a = u_0 + 2\alpha_1; \h{1cm}
u_c = u_0 + 2\alpha_2 .
\end{eqnarray}
allowing Eq. (\ref{eq-y-1}) to be rewritten as a cubic equation
\begin{eqnarray}
 \label{eq-cubic}
\nonumber
 y^3 +[u_0 \!-\alpha_1 \!-\alpha_2]y^2 + \!\left[ \frac{u_0^4}{4}-u_0 \left( \alpha_1 + \alpha_2 \right)
- \alpha_1^2 - \alpha_2^2 \right]y \nonumber \\
- \frac{u_0}{2}\left[ \frac{u_0}{2}\left( \alpha_1 + \alpha_2 \right)
+  \alpha_1^2 + \alpha_2^2 \right] =0,\h{2.7cm}\n
\end{eqnarray}
provided that $y \neq -\frac{u_0}{2}$, where an equality would lead to zero denominators in Eq.~(\ref{eq-y-1}). As can be easily checked, $y=-u_0/2$ is also a root of Eq. (\ref{eq-cubic}), which corresponds to the value of $y_b$ of the dark state $\psi_d$. Therefore the cubic equation can be reduced to a quadratic equation
\begin{eqnarray}
 \label{eq-quadratic}
 y^2 + \left(\frac{u_0}{2} - \alpha_1 -\alpha_2   \right)y
\nonumber -
 \left[\frac{u_0}{2}\left( \alpha_1 +\alpha_2 \right)
+ \alpha_1^2 + \alpha_2^2\right]  =0
\end{eqnarray}
with the roots
\begin{eqnarray}
 \label{eq-roots}
 y_{\pm} \!= \frac{\alpha_1\! + \alpha_2 }{2}-\frac{u_0}{4}
\!\pm
\sqrt{\left[\frac{u_0}{4}+\frac{ \alpha_1 \!+ \alpha_2 }{2} \right]^2\h{-3mm}+ \alpha_1^2 + \alpha_2^2},
\end{eqnarray}
where the value $y_+$ corresponds to $\psi_g$ and  $y_-$ to $\psi_e$.

\subsection{Choice of the well depths}

The ground and the excited states satisfying the optimal condition Eq.~(\ref{eq-ucua}) have now been determined in terms of the parameter $u_0$ which which remains a free parameter, and therefore allows for further optimization.  That will be based on the requirement that the process stays as close to adiabatic as possible, with negligible transition from the dark state to the other two available states. A measure of the transition rate is given by the parameter
\begin{eqnarray}
\label{eq-calA}
{\cal A}(\psi_i,\psi_d) = \frac{\left\langle \psi_i\left| \frac{\partial H}{\partial t} \right| \psi_d \right\rangle}{|\langle \psi_i|H| \psi_i \rangle - \langle \psi_d|H| \psi_d \rangle|^2}
\end{eqnarray}
where energy gap between dark state ($d$) and the adjacent states ($i=g,e$) sets the scale \cite{Cole}.

A rigorous optimization of $u_0$ with respect to (\ref{eq-calA}) can be done numerically, but our goal here is to get intuitive analytical expressions for the system parameters. Since a smaller value of ${\cal A}(\psi_i,\psi_d)$ is better in the sense of lesser transfer to the other states, we can deduce simple relations by requiring the denominators in (\ref{eq-calA}) for both $\psi_g$ and $\psi_e$ to be as large as possible. For ${\cal A}(\psi_i,\psi_d)$ to be small for both the states, the energy of the dark state should be exactly between the energies of the ground and excited states:
\begin{eqnarray}
\left(1+\frac{u_0}{2}\right)^2 = \frac{1}{2}
\left[\left(1-y_+\right)^2
+\left(1-y_-\right)^2 \right],
\end{eqnarray}
from which it follows that (up to the first order in $\alpha_{1,2}$)
\begin{eqnarray}
 u_0=-2(\alpha_1 + \alpha_2),
\end{eqnarray}
so that the optimal choices of potentials $u_a$ and $u_c$ are
\begin{eqnarray}
\label{eq-ua}
 u_{a} = -2\alpha_{2}\h{1cm} u_{c} = -2\alpha_{1}.
\end{eqnarray}
These relations show that to maintain minimum population in the central well, one needs to vary the well depths of the extreme wells in inverse proportion to the barrier heights. But, remarkably, the variation of the depth of each extreme well has to be synchronized with the height of the inter-well barrier which is \emph{non-adjacent} to it.

These conditions also fix $y_{\pm}$
\begin{eqnarray}
 y_{\pm}= \alpha_1 + \alpha_2 \pm \sqrt{ \alpha_1^2 + \alpha_2^2 },
\end{eqnarray}
which in turn determine the instantaneous energies of the three states:
\begin{subeqnarray}
\label{eq-e1}
 E_{g,e} &=& \frac{\hbar^2 k_0^2}{2m}\left[ 1-2 \left(\alpha_1 + \alpha_2 \pm \sqrt{ \alpha_1^2 + \alpha_2^2 } \right)\right] ,\\
E_d &=& \frac{\hbar^2 k_0^2}{2m}\left[ 1-2 \left(\alpha_1 + \alpha_2  \right)\right].
\end{subeqnarray}
as well as the parameters that determine the instantaneous forms of the ground and the excited states, respectively,
\begin{subeqnarray}
y_{a,c} &=& \alpha_{1,2} \pm \sqrt{ \alpha_1^2 + \alpha_2^2 }, \\
y_b &=& \alpha_1 +\alpha_2 \pm \sqrt{ \alpha_1^2 + \alpha_2^2 }, \\
\phi &=& \pm \frac{(\alpha_1 -\alpha_2)(\alpha_1 +\alpha_2 \pm \sqrt{ \alpha_1^2 + \alpha_2^2 })\pi}{2\sqrt{ \alpha_1^2 + \alpha_2^2 }} , \\
A,C &=& \mp \frac{\alpha_{1,2}}{\sqrt{2( \alpha_1^2 + \alpha_2^2) }}, \h{1cm}
B =\frac{ 1}{\sqrt{2}}.
\end{subeqnarray}
The $+$ corresponds to the ground state and $-$ to the excited state.  The knowledge of these states and their energies are essential for optimal time evolution.

\section{Finite width barriers}

We now show that the delta function potentials capture the behavior of high and narrow finite width inter-well barriers quite well, and the conclusions derived above remain qualitatively unchanged even with wider barriers. We relax the delta potentials to consider square barriers of width $D$, and height $U_{1,2}$ and as before introduce small dimensionless  parameters $\alpha_{1,2}$,  now defined to be
\begin{eqnarray}
U_{1,2} = \frac{\hbar^2k_0}{2\pi m D \alpha_{1,2}}.
\end{eqnarray}
We search for the conditions that would allow for the dark state of the sine form as in the previous case, requiring the wavefunction to have the form
\begin{eqnarray}
 \label{eq-psi-1fin}
 \psi_d(x)=\left\{
 \begin{array}{ll}
 A \sqrt{\frac{2}{L}}
 \cos \left[k_0(1-y_a)(x+D)- \frac{3\pi}{2}y_a \right], & x\in a \\ 
\\
 B\sqrt{\frac{2}{L}}\cosh \left[ \kappa_1 \left(x +\frac{\pi}{k_0} \right) \right] & x\in a'  \\
  B\sqrt{\frac{2}{L}}\sin  \left[k_0(1-y_b)x \right], & x\in b \\ 
 B\sqrt{\frac{2}{L}}\cosh \left[ \kappa_2 \left(x -\frac{\pi}{k_0} \right) \right] & x\in c'  \\
  C\sqrt{\frac{2}{L}}\cos \left[k_0(1-y_c)(x-D)+ \frac{3\pi}{2}y_c \right], & x\in c \\ 
\end{array}
 \right.\n
\end{eqnarray}
with
$a =[-\frac{3}{2}L-D ; -\frac{1}{2}L -D]$,$
a' =[-\frac{1}{2}L-D ; -\frac{1}{2}L ]$,
$b =[- \frac{1}{2}L ; \frac{1}{2}L]$,
$c'  =[ \frac{1}{2}L ; \frac{1}{2}L+D  ]$,
 $c  =[ \frac{1}{2}L+D ; \frac{3}{2}L+D]$.
The same coefficient $B$ is chosen for the barriers and the middle well, because
up to the second order the derivatives are zero at the boundaries. The equivalence of the energy of the stationary states in all the wells leads to a condition similar to Eq.~(\ref{eq-ucua}) for the delta barriers
\begin{eqnarray}
\label{eq-ucuafin}
 u_c-u_a = 2\left(y_c-\frac{y_c^2}{2}-y_a + \frac{y_a^2}{2}\right),
\end{eqnarray}
The only difference is that in order to allow for finite width, we retain second order terms in $y_{a,c}$, assuming that they are still relatively small.   The continuity conditions for the wavefunction and its derivative leads to the equations for  $y_{a,c}$:
\begin{eqnarray}
\label{kappak}
 \kappa_{1,2} \tanh \left( \kappa_{1,2} D\right) = -\frac{k_0(1-y_{a,c})}{\tan[\pi (1-y_{a,c})]},\n\\
{\rm with}\ \  \frac{\kappa_{1,2}}{k_0}=\sqrt{[{\pi k_0 D \alpha_{1,2}}]^{-1}-1}.
\end{eqnarray}
For narrow ($k_0D \ll 1$) and high ($U_{1,2}\gg \hbar^2 k_0^2/2m$) barriers this reproduces Eq.~(\ref{eq-yaq}) for delta barriers,  $y_{a,c}\simeq  \alpha_{1,2}$.
Analytical expressions for $y_{a,c}$ accurate to higher order can be easily obtained from the equation, which can then used to get the well depths in Eq. (\ref{eq-ucuafin}).

Likewise the amplitude of the particle to be inside the middle well is
\begin{eqnarray}
  \frac{b}{a}=\frac{k_0}{\kappa_{1} \sinh \left( \kappa_{1} D\right)}.
\end{eqnarray}
which, for a narrow barrier $\kappa_{1} D\ll 1$, also reduces to
$b\approx a\pi \alpha_1$, the result of the delta barriers.

Although the specific relations may vary with the size and shape of the  potentials, the essential fact remains that the well depths need to be varied in sync with the barrier heights to maintain negligible population in the central well.  It is interesting to note that by increasing the barrier widths and lowering its magnitude, keeping $\alpha_{1,2}$ constant, the probability of the particle to be in the middle well can be reduced further than for delta barriers, since the $\sinh$ function increases faster than linearly for larger $\kappa_{1} D$.

\begin{figure}[b]
\centering
\includegraphics[width=\columnwidth]{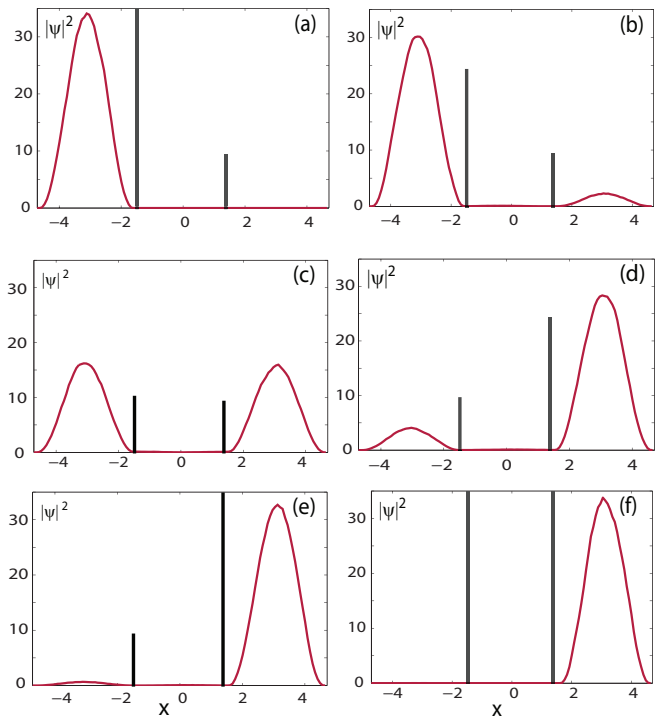}
\caption{Numerical simulation of the time dependent probability density for the particle confined to three wells. The parameter $1/\alpha_{1,2}$ vary between 30 and 1550 as shown in Fig. \ref{fig-qs}. The plots correspond to times $(a)$ $t=1400\tau$, $(b)$ $t=4,900\tau$, $(c)$ $t=5,880\tau$, $(d)$ $t=7,000\tau$,
$(e)$ $t=7,700\tau$, and $(f)$ $t=14,000\tau$  with $\tau=mL^2/(\pi^2 \hbar)$ .
\label{fig-dens}}
\end{figure}

\begin{figure}[b]
\centering
\includegraphics[width=0.8\linewidth]{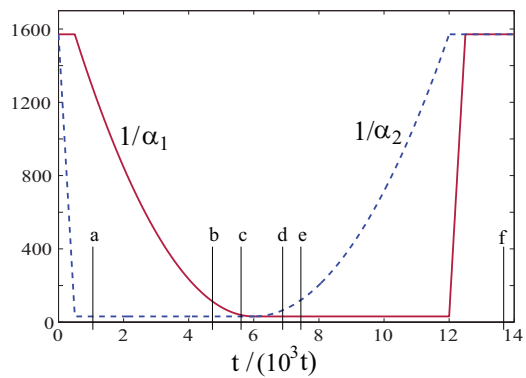}
\caption{Time evolution of the parameters $1/\alpha_{1,2}$ characterizing the magnitude of the delta potentials (see Eq. (\ref{eq-ualpha})) with the time scale $\tau=mL^2/(\pi^2 \hbar)$. The marks $(a)$ through $(f)$ correspond to the snapshots in Fig. \ref{fig-dens}.
\label{fig-qs}}
\end{figure}

\section{Transport procedure}

We apply the conditions derived above to demonstrate optimal transfer.
We start with a particle in the ground state of well $a$ which is isolated by a very large barrier from the neighboring well $b$ (i.e., ${\cal U}_1\to \infty$, $\alpha_1 \approx0 $).
\begin{enumerate}
\item
The barrier between $b$ and $c$ is lowered, i.e., $\alpha_2$ reaches a finite (but small) value whereas $\alpha_1 \approx0 $. At the same time the potential of well $a$ is lowered to $u_a=-2/\alpha_2$, according to Eq. (\ref{eq-ua}). The state of the particle then coincides with the dark state having $A_d\approx -1$, and $B_d,C_d\approx 0$. The two remaining states   form superpositions of the states occupying wells $b$ and $c$: $B\approx C\approx 1/\sqrt{2}$ (ground state), and  $B\approx -C\approx 1/\sqrt{2}$ (excited state). The energies of these two states are separated by $\approx \hbar^2k_0^2\alpha_2/ m$ from the energy of the middle state. So far the particle stays in well $a$.
\item
The barrier between $a$ and $b$ is gradually lowered to the same value as that between $b$ and $c$, $\alpha_1=\alpha_2$, with simultaneous lowering of the potential in well $c$, $u_c=u_a=-2/\alpha_2$.
The state of the particle adiabatically passes to the superposition of occupying the wells $a$ and $c$ with the same probability. At the end of this step the particle is in the state with coefficients
$A_d\approx -C_d \approx 1/\sqrt{2}$, the remaining two states being separated from it by the energy of $\approx \sqrt{2}\hbar^2k_0^2\alpha_{1,2}/m$. The probability of occupying the middle well $b$ is in this stage $\approx \pi^2 \alpha_{1,2}^2/2$.
\item
The barrier between $b$ and $c$ is gradually increased ($\alpha_2\to 0$) simultaneously bringing the potential energy of well $a$ to $u_a=0$. The middle state adiabatically goes to the state localized in well $c$, with coefficients $A_d,B_d \approx 0$, $C_d\approx 1$. The remaining two states now form superpositions with $A\approx B\approx 1/\sqrt{2}$ (ground state), and  $A\approx -B\approx -1/\sqrt{2}$ (excited state). The energies of these two states are separated by $\approx \hbar^2k_0^2\alpha_1/ m$ from the energy of the middle state.
\item
The barrier between $a$ and $b$ is increased to its original value, $\alpha_1\to 0$ while bringing the potential energy of well $c$ back to $u_c=0$. Nothing happens to the particle which is now localized in well $c$.
\end{enumerate}
This procedure was implemented by a direct numerical solution of
the time dependent Schr\"{o}dinger equation by a split-step operator method, where the time propagation is done alternately in incremental steps in position and momentum spaces, transforming between the two via Fast Fourier Transforms.  Narrow Gaussian potentials were used instead of delta functions, based upon the considerations of the previous section. Snapshots of the evolution are shown in Fig.~\ref{fig-dens} and the corresponding variation of the barrier heights shown in  Fig.~\ref{fig-qs}.  The simulation shows a smooth transition of the wavepacket from well $a$ into well $c$, with essentially no population in the central-well at any time showing the effectiveness of the optimal conditions we derived.

\section{Current in the Central Well}

When the potential barriers are changing at a finite rate the wavefunction differs from the eigenstate of the instantaneous Hamiltonian. This change influences the resulting probability of occupying the middle well.  The middle well occupancy and associated current can be estimated using the continuity equation
\begin{eqnarray}
 \frac{\partial \varrho}{\partial t}=-\frac{\partial j}{\partial x}; \h{1cm}j=\frac{i\hbar}{2m}\left( \psi \frac{\partial \psi^*}{\partial x}-
 \psi^* \frac{\partial \psi}{\partial x} \right).
\end{eqnarray}
relating the probability density $\varrho = |\psi|^2$  to the current density $j$. Integrating the probability density in wells $a$ and $c$ and assuming no accumulation of probability in well $b$ it follows for the current $j_b$ through well $b$ to be
\begin{eqnarray}
 j_b=  \frac{d|c|^2}{dt}=- \frac{d|a|^2}{dt}=
\frac{2\alpha_1 \alpha_2 (\alpha_2\dot{\alpha_1}-\alpha_1\dot{\alpha_2})}{(\alpha_1^2+\alpha_2^2)^2},
\end{eqnarray}
where the dot denotes time derivative.
To allow for such a current in well $b$ the sine-form wavefunction has to be complemented with an imaginary cosine part such that
\begin{eqnarray}
 \psi_d \approx b \sqrt{\frac{2}{L}}\sin (k_0 x)+i b_2\sqrt{\frac{2}{L}}\cos (k_0 x),
\end{eqnarray}
where
\begin{eqnarray}
 b_2=\frac{mL^2}{\pi^2\hbar }\frac{(\alpha_2\dot{\alpha_1}-\alpha_1\dot{\alpha_2})}{(\alpha_1^2+\alpha_2^2)^{3/2}} .
\end{eqnarray}
A dimensionless time derivative $\alpha_{1,2}'\equiv \tau \dot{\alpha}_{1,2}$ is defined in terms of the time scale $\tau\equiv mL^2/(\pi^2 \hbar)$, which is of the order of the round-trip time $T=4mL^2/(3\pi \hbar)$ of the particle between the walls of an infinite well of length $L$:
\begin{eqnarray}
 b_2= \frac{\alpha_2\alpha'_1-\alpha_1 \alpha'_2}{(\alpha_1^2+\alpha_2^2)^{3/2}} .
\end{eqnarray}
This value should satisfy the condition $b_2\ll 1$. In the case of $\alpha_1 \approx \alpha_2$ and $\alpha'_1 \approx -\alpha'_2$ this leads to the requirement $\alpha'_{1,2}\ll \alpha_{1,2}^2$. The dynamical contribution to the population of well $b$ is less than the static one, $|b_2|^2<|b|^2$, provided that $\alpha'_{1,2}\lesssim \pi \alpha_{1,2}^3$. Even though for different time dependencies of $\alpha_{1,2}(t)$ we get different probabilities of occupying the middle well, we can use these results to estimate the trade-off between the highest probability $p_b$ of being in well $b$ and the time ${\cal T}$ necessary to complete the transport between $a$ and $c$, namely $p_b{\cal T}\gtrsim \pi \tau$.

It is interesting to interpret the vanishing probability in the central well $b$ in terms of the particle velocity. Classically, the current density is related to density and mean velocity by $j=\rho v$, from which the classical velocity in the middle of well $b$ would be (assuming the simple case  $\alpha_1 \approx \alpha_2$ and $\alpha'_1 \approx -\alpha'_2$)
\begin{eqnarray}
v=\frac{j_b}{\rho_b}\approx \frac{L}{\tau}\frac{\alpha_{1,2}^3}{\alpha'_{1,2}}
=\pi v_0\frac{\alpha_{1,2}^3}{\alpha'_{1,2}},
\end{eqnarray}
where $v_0= \hbar k_0/m$ is the ground state velocity. Thus for
$\alpha'_{1,2}\ll \alpha_{1,2}^3$ (i.e., when the dynamical contribution to the middle-well probability is much less than the static one) the negligible probability of the particle to be in the central well can be understood as arising from the particle speeding through the middle well much faster than the velocity it has inside wells $a$ and $c$.

\section{Conclusions}

We presented analytical results for coherent tunneling via adiabatic passage in a system of three square wells separated by barriers that are high and narrow compared to the well dimensions. We have shown that in order to maintain negligible population in the central well, required by the CTAP process, the depths of the extreme wells need to be varied as well as the barrier heights; this has the counterintuitive behavior that, during the time evolution, the depth of each exterior well needs to be correlated with the height of the barrier non-adjacent to it. We determined those correlations along with expressions for the relevant stationary states and their energies satisfying them, which we apply in a numerical solution of the time dependent Schr\"{o}dinger equation to demonstrate that they indeed lead to transfer between the extreme wells with virtually no occupation of the central well.

Our general conclusion, that simultaneous and synchronized variation of both well depths and barrier heights is essential for optimal transfer, applies qualitatively to more general potentials, although the exact nature of the relations may be different.  For finite rate of time variation we found a relation for the tradeoff between central well occupancy and the transfer rate. We also provide a novel interpretation of the transfer mechanism based upon a current through the central well.

The results here can find applications in coherent transport of electrons between quantum dots or atoms in micromagnetic traps, where our expressions can serve as a guide for choosing parameters for the CTAP process.

\acknowledgments
T.O. is supported by the Czech Institutional research plan MSM6198959213. K. D. acknowledges support of the Research Corporation in the initial stages of the work.

\end{document}